# Mechanical properties of Nanotubes of Polyelectrolyte Multilayers

*Stephane Cuenot[1]\*, Halima Alem[2], Guy Louarn[1], Sophie Demoustier-Champagne[2], Alain M. Jonas[2].*

[1]Institut des Matériaux Jean Rouxel (IMN), Université de Nantes, 2, Rue de la Houssinière, 44322 Nantes cedex 3, France.

[2]Unité de Physique et de Chimie des Hauts Polymères, Université catholique de Louvain, Place Croix du Sud, 1, B-1348 Louvain-la-Neuve, Belgium.

**List of the authors:**

**Stéphane Cuenot** (corresponding author):

[1]Institut des Matériaux Jean Rouxel (IMN), Université de Nantes, 2, Rue de la Houssinière, 44322 Nantes cedex 3, France.

Stephane.Cuenot@cnrs-imn.fr

++33 240376421 (Phone number)

++33 240373991 (Fax number)

**Halima Alem :**

[2]Unité de Physique et de Chimie des Hauts Polymères, Université catholique de Louvain, Place Croix du Sud, 1, B-1348 Louvain-la-Neuve, Belgium.

Halima.Alem@mines.inpl-nancy.fr

 ++33 383584334 (Phone Number)

++33 383584344 (Fax Number)

**Guy Louarn :**

[1]Institut des Matériaux Jean Rouxel (IMN), Université de Nantes, 2, Rue de la Houssinière, 44322 Nantes cedex 3, France.

Guy.Louarn@cnrs-imn.fr

++33 240376416 (Phone number)

++33 240373991 (Fax number)

**Sophie Demoustier-Champagne** :

[2]Unité de Physique et de Chimie des Hauts Polymères, Université catholique de Louvain, Place Croix du Sud, 1, B-1348 Louvain-la-Neuve, Belgium.

demoustier@poly.ucl.ac.be

++32 10472702 (Phone Number)

++32 10451593 (Fax Number)

**Alain Jonas** :

[2]Unité de Physique et de Chimie des Hauts Polymères, Université catholique de Louvain, Place Croix du Sud, 1, B-1348 Louvain-la-Neuve, Belgium.

alain.jonas@uclouvain.be

++32 10473765 (Phone Number)

++32 10451593 (Fax Number).

# Mechanical properties of Nanotubes of Polyelectrolyte Multilayers

*Stephane Cuenot[1]\*, Halima Alem[2], Guy Louarn[1], Sophie Demoustier-Champagne[2], Alain M. Jonas[2].*

[1]Institut des Matériaux Jean Rouxel (IMN), Université de Nantes, 2, Rue de la Houssinière, 44322 Nantes cedex 3, France.

[2]Unité de Physique et de Chimie des Hauts Polymères, Université catholique de Louvain, Place Croix du Sud, 1, B-1348 Louvain-la-Neuve, Belgium.

\* corresponding author

PACS number: 62.25.Jk

Abstract:

The elastic properties of nanotubes fabricated by layer-by-layer (LbL) assembly of polyelectrolytes in the nanopores of polycarbonate track-etched membranes have been investigated by resonant contact Atomic Force Microscopy (AFM), for nanotube diameters in the range of 100 to 200 nm. The elastic modulus of the nanotubes was computed from the resonance frequencies of a cantilever resting on freely-suspended LbL nanotubes. An average value of 115 MPa was found in air for the Young's modulus of these nanostructures, well below the values reported for dry, flat multilayers, but in the range of values reported for water-swollen flat multilayers. These low values are most probably due to the lower degree of ionic cross-linking of LbL nanotubes and their consequently higher water content in air, resulting from the peculiar mode of growth of nano-confined polyelectrolyte multilayers.

KEYWORDS: TEM, AFM, nanotubes, polyelectrolyte, mechanical properties, surface effects.

# 1. Introduction :

Layer-by-layer (LbL) assembly of polyelectrolytes was recently demonstrated within nanoporous templates such as track-etched or alumina membranes: LbL is performed by repeatedly passing through the membranes solutions of polyelectrolytes of opposite charge [1-7]. After dissolving the templating membrane, LbL nanotubes and nanowires are readily obtained, which could be of interest for applications such as drug encapsulation, gene delivery, or sensing. In a previous report [7], it was shown that LbL nanotubes do not grow as regular, standard polyelectrolyte multilayers. Indeed, polyelectrolyte chains entangle during their passing through the nanopores of membranes, due to a local increase of concentration, therefore making a relatively dense gel filling the most part of the nanopores. For flat, standard multilayers, multilayer growth involves the successive surface adsorption and surface formation of a polyelectrolyte complex, with diffusion into previously adsorbed layer(s) only occurring for a few specific systems [8]. In the case of LbL nanotubes, however, tube formation results from the formation of a polyelectrolyte complex by the successive percolation of polyelectrolytes through the swollen gel resulting from the entanglement in the nanopores of previously filtered polyelectrolyte chains. Upon drying, the swollen complex collapses on the inner surfaces of the nanopores, resulting in nanotubes of wall thickness almost proportional to the radius of the nanopores and in any case much larger than the thickness obtained for comparable flat multilayers. The mode of formation of LbL assemblies inside nanopores is thus very different from standard LbL multilayers, which implies that their structure and properties should be significantly different from those of usual LbL assemblies.

Here, we investigate the mechanical properties of such LbL nanotubes, and compare them with those of standard, flat LbL multilayers. The mechanical properties of standard LbL multilayers have been studied for quite some time (Table 1) [9-27]. In the dry state, Young moduli in the 1-13 GPa range were reported for various systems using a variety of testing methods, typical moduli being about 5 GPa for purely organic assemblies, and rising to about 10 GPa upon addition of inorganic components. These values are classical for amorphous polymers and composites below their glass transition temperature. In water, the range of published moduli is lower, and depends on details such as ageing, ionic strength, temperature, template nature (for micro-capsules), persistence length of the polymers, etc. Most authors report values between 1 and 600 MPa, although Picart et al. [23] reported values as low as 20 kPa for a specific system. These are usual values for elastomers, indicating that water plasticizes the multilayers. However, the mechanical properties of LbL nanotubes have not been measured so far, probably because the nanometer-range diameter of LbL nanotubes, and their lengths limited to a few micrometers, conspire to make their mechanical testing especially challenging. Therefore, to probe the mechanical properties of these tiny objects, we have adapted a resonant AFM method previously used to study the mechanical properties of stiffer nanotubes or nanowires of metals and conducting polymers [28-30]. We present the results obtained on several nanotubes, and discuss their Young's modulus with respect to the peculiarities of their formation.

## 2. Materials and Method:

### 2.1 Materials:

A poly(vinylbenzylammonium chloride) (polycationic PVBAC) sample (Mw=17800 g/mol; polydispersity=1.24) was synthesized from poly(vinylbenzylchloride) (PVBC, Polymer Source, Canada) and purified as described previously [7]. A poly(styrene sodium sulfonate) (polyanionic PSS) sample (Mw=13800 g/mol; polydispersity= 1.09) was obtained  by the sulfonation of a PS sample (Polymer Source, Canada) as described in [7]. The structure and characterization of these polyions was fully reported in [7]. Polycarbonate track-etched membrane (PC) (103 and 194 nm pore diameter) were supplied by It4ip, Seneffe, Belgium. The Aluminum oxide ($Al_2O_3$, Aldrich) and dichloromethane ($CH_2Cl_2$, spectrometric grade, Acros) were used without any further purification. The water used in all experiments was purified with a Millipore system to a resistivity of 18.2 $M\Omega$.cm (Milli-Q water).

*2.2 Preparation of LbL nanotubes :*

Nanotubes were obtained by LbL deposition in the nanopores of PC track-etched membranes of 20 μm thickness, starting from PSS and PVBAC solutions $10^{-2}$ mol of monomer per liter of water (no salt added). Briefly, the PVBAC solution was first filtered under a pressure of 4 bar through the membrane. The top surface of the membrane was then washed with pure water to remove the cake of unfiltered material, and pure water was filtered three times through the membrane under a pressure of 4 bar, to rinse the pores. The PSS solution was then filtered similarly, thereby complexing both polyelectrolytes. This procedure was performed twice, giving rise to produce (PVBAC/PSS)$_2$ nanotubes. The dimensions of the nanotubes before extraction were determined by Transmission Electron Microscopy (TEM) as described in [7]. After LbL assembly, the surface of the membrane was polished by powdered $Al_2O_3$ to remove the polyeletrolyte complex formed at the top surface of the membrane; the membrane was then rinsed in Milli-Q water and dissolved in $CH_2Cl_2$ to extract the nanotubes. Finally, drops of this suspension were deposited on carbon grids for TEM observations, or on poly(ethylene terephthalate) track-etched membranes for AFM imaging and mechanical measurements. After deposition, the nanotubes were rinsed a few times with pure $CH_2Cl_2$ to remove any residual trace of polycarbonate.

*2.3 Characterization of extracted nanotubes by AFM:*

AFM was used to image the nanotubes and to measure their mechanical properties. All experiments were performed with an Autoprobe® CP microscope (Thermomicroscopes) operated in air with 100 μm or 5 μm scanners, equipped with ScanMaster® detectors correcting for drift, non-linearity and hysteresis effects.

*2.3.1 Imaging:*

The images were recorded in contact mode (topography, C-AFM). The cantilevers were silicon standard $Si_3N_4$ Microlevers$^{TM}$ with integrated pyramidal tips (typical apex radius of curvature between 30 and 50 nm). Only a second order flattening procedure (line by line) was performed on AFM images.

*2.3.2. Mechanical properties:*

The nanotube suspension was deposited on poly(ethylene terephthalate) (PET) membranes with pore diameters of about 1 µm. This suspension was not filtered through the pores because the applied pressure would make the soft nanotubes plunge within the pores. In order to remove any contaminant from the nanotube surfaces, especially traces of polycarbonate, the samples were carefully rinsed with pure dichloromethane. The measurement of the mechanical properties of LbL nanotubes suspended over pores was then performed by AFM in the electrostatic resonant contact mode [28]. In this mode, an alternative external electric field is applied between the sample holder and the microscope head and induces the cantilever vibration. The polarization forces acting on the tip induce a deflection of the cantilever. By varying the frequency of the electric field, it is possible to completely characterize the resonance spectrum of the cantilever while the tip contacts or not the surface of the suspended nanotube. The mechanical properties of the nanotube can then be obtained from this spectrum. Note that, in bending tests such as performed here, the beam deflection induces both tensile/compressive deformations and shear deformations. To minimize the shear component to less than 10% of the whole contribution, nanostructures for which the ratio between suspended length and height is larger than 10 were only measured. For mechanical measurements, the cantilevers were the same as those used for the imaging. The spring constant of each cantilever was of 0.1 N/m. Geometrical characterization of the cantilever was realized by high-resolution scanning electron microscopy. Obtained data were used for the description of the dynamical behavior of the cantilever using the Rayleigh-Ritz approximation [31-32]. In the Rayleigh-Ritz procedure, modal analysis of the cantilever results from a variational principle applied to the Rayleigh quotient between the potential and kinetic energy of the cantilever. By this procedure, the resonance frequencies can be calculated without having to solve the differential equation of motion of the cantilever.

The physical properties of the cantilever material (i.e., its elastic modulus and its density) were deduced from the free experimental resonance frequency of the cantilever. The modulated electric field was applied between the sample holder and the AFM head using a function generator (Agilent Technologies, model 33120A). The cantilever deflection signal was measured using a lock-in amplifier (EG&G Princeton Applied Research, model S302). The signal generator command and the data collection from the lock-in were computerized and data analysis was realized using routines developed under Igor Pro software (Wavemetrics).

## 3. Results and discussion:

Selected TEM micrographs of extracted nanotubes (Figure 1) show that the shape of the nanostructures is modified when they are deposited on surfaces. The initial perfectly cylindrical shape of the nanotubes within the membrane pores, which was checked in a previous report by TEM observations of microtomed membranes after LbL deposition [7], is frequently replaced by a flattened shape when the nanotubes are extracted from the membrane and brought in contact with a surface. This flattening arises from the competition between the elastic energy due to stretching, bending and shear deformations on the one hand, and the interaction energy resulting from forces between the walls of the tubes or between the tubes and the surface of the membrane, on the other hand. In one-dimensional systems such as nanotubes, the stretching and shear deformation modes are energetically expensive relative to the bending modes. As a consequence, such systems can be considered as inextensible, and the obtained shape essentially results from the competition between bending elasticity on the one hand, and adhesion or wall-to-wall interaction on the other hand. Simple approximations of van der Waals forces indicate that the walls of the tubes tend to attract each other [33]. However, the strength of the wall-to-wall interaction is strongly dependent on the medium into which the tubes are immersed, since the Hamaker constant related to the van der Waals interaction of the walls is very different in air or in a solvent such as methylene chloride or water. Because the indexes of refraction of water and methylene chloride are much closer to the ones of polymer materials than the one of air, collapse of the tubes due to van der Waals forces is more probable to occur in air than in methylene chloride or water [34]. Other factors may also modulate the collapse of the tubes, such as the electrostatic repulsion of the charged walls in water, the entropic retraction force of possible tie molecules connecting opposite walls, or the adhesion force with a surface when the tubes are adsorbed on the membrane. It is difficult to predict the balance of these forces in a given medium. When the tubes are adsorbed and imaged in air, however, our results clearly indicate that surface energy and wall-to-wall interaction energy dominate over the elastic bending energy, which results in both axial and radial deformations of the nanotubes. In a previous work, the cylindrical shape of polypyrrole (PPy) nanotubes having an elastic modulus larger than 1GPa, was found to be kept after nanotube dispersion onto similar surfaces [28-30]. This suggests that the elastic modulus of LbL nanotubes is somewhat lower. In the sequel, we will refer to undeformed polyelectrolyte nanotubes as "LbL nanotubes" (i.e., before extraction from the membrane or when in suspension), and to surface-flattened nanotubes as "LbL nanostructures".

The nanotubes were then dispersed on PET track-etched membranes, rinsed, and imaged by AFM. Large-scale images (typically ~10x10 $\mu m^2$) were acquired to select nanostructures suspended over pores that could be used to measure their mechanical properties. Once a suspended nanostructure was located, its image was then realized at a lower scale to determine its precise dimensions, i.e., its suspended length, L, and its height, $h_{AFM}$ (Figure 2).

The mechanical properties of such suspended nanostructures were then assessed by AFM in the electrostatic resonant contact mode, as described in the Experimental Section. The method involves measuring the resonance spectrum of the AFM cantilever when the tip is in contact or not with the suspended nanostructure (Figure 3), from which mechanical properties of the nanostructure can be obtained [28]. Therefore, the AFM tip was positioned midway along the suspended length of the nanostructure and the resonance spectrum of the cantilever was measured. A typical spectrum obtained on a LbL nanostructure is presented in Figure 4 (the first resonance peak of the free cantilever is shown in inset, for comparison).

Three peaks are observed in the spectrum of the cantilever-tip system in contact with the nanostructure. Like macroscopic beams, cantilevers can vibrate in different types of modes, such as flexural, torsional or longitudinal modes. The fundamental vibration mode of a cantilever, i.e., the one with the lowest resonance frequency, always corresponds to a flexural vibration mode. In Figure 4, the two stronger ones ($F_1$, $F_2$ in Fig.4) correspond to flexural cantilever vibrations, whereas the smaller one ($T_1$ in Fig.4) results from torsional vibrations. This was deduced from the relative contributions of each resonance peak to the vertical and lateral signals provided by the position-sensitive photodiodes of the AFM (couplings in the horizontal and vertical detection systems were responsible for partial signal mixing). The first resonance frequency of the cantilever increases from 64 kHz when the tip is not in contact with the nanostructure to 120 kHz when the tip is brought into contact. As expected, by changing the boundary conditions at the tip apex, the first resonance frequency of the cantilever in contact with the nanostructure was higher than the corresponding free resonance frequency. This frequency shift reflects the dynamics of the global system made of the cantilever, tip, and nanostructure. The compliance of this global system is due to the bending of the nanostructure itself and to the mechanical deformation at the tip-sample contact, which is much smaller than the former [28] and will be neglected in the sequel.

Previously, a similar method was used to determine the elastic properties of metallic nanowires and of nanotubes of conjugated polymers [28-30]. In both cases, the mechanical behavior of the cantilever-tip-nanotube system could be modeled properly as the one of a cantilever in contact with two springs of identical stiffness $k_{nanost}$, representing the vertical and lateral stiffness associated to the deformation of the nanostructure in the two axial directions, respectively [28].

Here, we assumed that a similar spring model holds for LbL nanostructures, and we neglected the inertial contribution of the nanostructure since the natural resonance frequency of the nanostructures is always much higher than that of the cantilever. Within this model, the resonance frequencies of the cantilever in contact with the nanostructure can be computed by the Rayleigh-Ritz method, provided the geometrical dimensions of the cantilever and $k_{nanost}$ are known. The cantilever dimensions were determined by high resolution scanning electron microscopy, and $k_{nanost}$ was adjusted to reproduce the experimental values of resonance frequency.

Once $k_{nanost}$ is known, the Young's modulus E of the nanostructures can be obtained, provided appropriate boundary conditions are taken. In the present case, a strong adhesion of the nanostructure to the PET membrane is supported by the observation that the nanostructures are not displaced upon extensive AFM imaging, whatever the relative orientation between the axis of the nanostructure and the fast scan direction of the AFM. Therefore, it may be safely considered that the nanostructures behave as clamped-beams with no possible vertical motion of their extremities in contact with the PET membrane. Under these circumstances [28],

$$\frac{1}{k_{nanost}} = \frac{L^3}{192EI} + \frac{1}{2k_s}$$

where I is the moment of inertia of the nanostructure, L is its suspended length, and $k_s$ is the stiffness of the nanostructure-on-membrane contact (this formula is valid when the tip is placed in the middle of the suspended length of the nanostructure). The nanostructure-on-membrane contact stiffness $k_s$ was determined so that the predicted frequency at the nanostructure ends corresponds to the frequency measured on the nanostructure lying on the membrane. The moment of inertia was estimated as described in the supporting information, taking into account the shape of the nanostructure and its measured dimensions.

The histogram of Young's moduli, obtained from 32 independent measurements, is presented in Figure 5. The elastic modulus values are comprised between 40 MPa and 225 MPa, and are distributed about an average value of 115, with a standard deviation of 50 MPa. For materials with nanometer length scales, due to the increasing surface-to-volume ratio with respect to macroscopic materials, surface effects may become predominant. In particular, in the present case where the nanostructure deformation induces an increase of its area, the surface tension effects have to be investigated. From our previous work, it is possible to derive a ratio that may be used to predict the onset of the surface tension effects [30]. More precisely, this ratio balances the stiffness of the suspended nanostructures due to the elastic modulus or due to surface tension. For the probed nanostructures, this ratio between the surface stiffness, $\kappa s$, and the nanostructure elastic stiffness, $\kappa t$, can be expressed as

$$\frac{\kappa s}{\kappa t} = \frac{1}{40}\frac{\gamma}{E}\frac{L^2}{I}\Phi(1-\nu)$$

where $\gamma$ is the surface tension of the material, $E$ is the material elastic modulus, $L$ is the length of the suspended nanostructure, $I$ is the moment of inertia of the nanostructure section, $\Phi$ is the contour length of the nanostructure section and $\nu$ is the Poisson's ratio.

This equation is established for a specific geometry of solicitation: suspended nanostructure with clamped-ends and a central solicitation [30]. When this ratio is larger than 1, surface tension effects prevail. For the probed nanostructures, the values of this ratio are comprised between 0.03 and 0.09, by using typical values for polymers of 0.4 for the Poisson's ratio and of 40 mJ m$^{-2}$ for the surface tension [30,35]. Therefore, although surface effects exist, they can be considered as a second-order correction. However, due to the dependence of this ratio on the geometrical dimensions of the probed nanostructures, surface effects could become more important for the same nanostructures synthesized within pores with lower diameters [30].

The obtained values of the elastic modulus are one to two orders of magnitude lower than the moduli previously determined on dry polyelectrolyte multilayers, and are actually in the range of values reported for multilayers in water (Fig.5 and Table 1). This indicates that LbL nanotubes are much softer in air than normal multilayers, in good agreement with their trend to flatten upon adsorption. As stated in the Introduction, the mode of growth of LbL multilayers in the nanopores of track-etched membranes differs strongly from the case of flat multilayers, with much larger increments of thickness per cycle of deposition, and complex formation occurring in a strongly entangled state resulting from the increased polyelectrolyte concentration in the pores [7]. These differences should translate into significantly different inner structures of the multilayers when adsorbed on flat surfaces or in nanopores. The present measurements confirm this expectation: the lower moduli measured on the nanotubes are in good agreement with a more swellable gel-like structure of the walls of LbL nanotubes, which would not be fully dried in air and would correspond to a lower degree of ionic cross-linking than normal, flat multilayers. The mechanical properties of the nanotubes are thus consistently lying in the range of those measured for water-swollen flat multilayers, even when the nanotubes are dried in air. This observation also indicates that the mechanical properties of polyelectrolyte multilayers crucially depend on the details of their growth mechanisms, which may differ from system to system, and on the geometry of their template (colloids, flat surfaces, or pores). This is most probably the reason for the large range of values reported in the literature for the moduli of polyelectrolyte multilayers (Table 1).

## 4. Conclusion:

The elastic properties of LbL-assembled nanotubes (with initial outer diameters between 100 and 200 nm), obtained by filtering solutions of polycations and of polyanions through the nanopores of polycarbonate track-etched membranes, have been investigated by resonant contact AFM. By monitoring the resonance frequency of a cantilever resting on freely-suspended LbL nanotubes, the elastic modulus of the nanotubes could be computed, as was done previously for nanotubes of metals or of organic conductors. An average value of about 115 MPa was found in air for the Young's modulus of these nanostructures, with a relatively narrow standard deviation of 50 MPa. This value is well below the reported values for elastic moduli of dry, flat multilayers, but lies in the range of values reported for water-swollen multilayers. This low value indicates that the LbL nanotubes are water-swollen in air, most probably due to their more loosely complexed structure resulting from their peculiar mode of growth outlined previously [7]. Our observations indicate that the elastic moduli of polyelectroyte multilayers crucially depend on the details of their formation, which may explain the dispersion of moduli values found in the literature.

**Acknowledgments :**

S.C. and G.L acknowledge the "Agence Nationale de la Recherche" for financial support in the frame of "Jeunes Chercheurs" program. Financial support by Belgian Federal Public Planning Service Policy (Inter-University Attaction Pole FS2), and the Fund for Scientific Research F.R.S.-FNRS is gratefully acknowledged. S.D.-C. is a Research Associate of the F.R.S.-FNRS.

**Table 1**. Literature results on the mechanical properties (elastic modulus, E) of LbL multilayers systems in water.

(1) PAH: poly(allyl amine hydrochloride); PSS: poly(styrene sulfonate); PAA: poly(acrylic acid); PDDA: poly(di-allyl di-methyl ammonium chloride); PLL: poly(L-lysine).

| LbL system (1) | E (GPa) | Testing method / Comments | Ref. |
|---|---|---|---|
| PAA/azobenzene | $10^{-4}$-$10^{-2}$ | AFM force-distance measurements / azobenzene containing polyelectrolytes | 14 |
| PAH/PSS | $10^{-3}$-$10^{-1}$ | AFM load-deformation on LbL micro-capsules / E depends on template, ageing, previous swelling and ionic strength | 16-19 |
| PSS/dendrimer/PAH | 0.08-0.15 | AFM load-deformation on LbL micro-capsules | 20 |
| DNA/PAH | 0.01-1 | AFM load-deformation on LbL micro-capsules | 21 |
| PLL/hyaluronan | $2x10^{-5}$ | AFM force-distance measurements / E was $8x10^{-4}$ when crosslinked | 22 |
| PDDA/PSS | 0.1 | AFM force-distance measurements on LbL micro-capsules / E decreases with T | 23 |
| PAH/PSS | 0.4 | AFM force-distance measurements on LbL micro-capsules / E decreases with ionic strength | 24 |
| carrageenan/PAH | $7.5x10^{-3}$ -$3.5x10^{-2}$ | AFM force-distance measurements / E depends on persistence length | 25 |

**Figures caption:**

**Figure 1.** (a) and (b) TEM images of (PVBAC/PSS)$_2$ nanotubes grown in a membrane of 103 nm average pore diameter. The nanotubes are flattened due to their interaction with the carbon-coated substrate.

**Figure 2.** C-AFM image of a (PVBAC/PSS)$_2$ LbL nanostructure suspended over the pore of a PET track-etched membrane.

**Figure 3.** Schematic drawing of our experimental setup showing the AFM tip located on the nanostructure suspended over the pores.

**Figure 4.** Typical resonance spectrum measured for a cantilever in contact with a suspended LbL nanostructure (logarithmic ordinate axis). The inset shows the first resonance peak of the free cantilever, for comparison.

**Figure 5.** Histogram of the moduli of 32 LbL nanostructures measured by the resonant AFM method. The range of values reported in the literature for the moduli of flat multilayers is also given in the figure, for measurements performed in water or in the dry state (data taken from Table 1).

**Figure 1.**

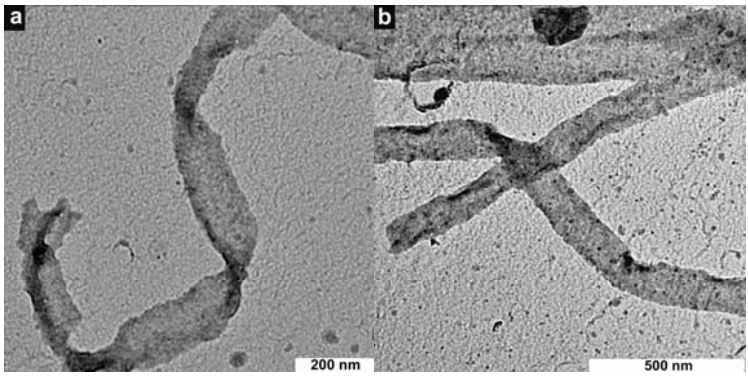

**Figure 2.**

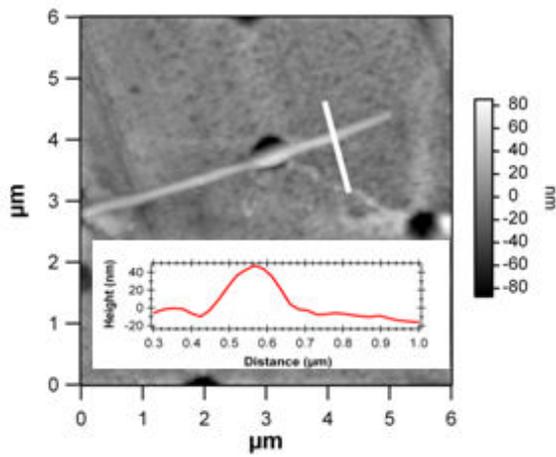

**Figure 3.**

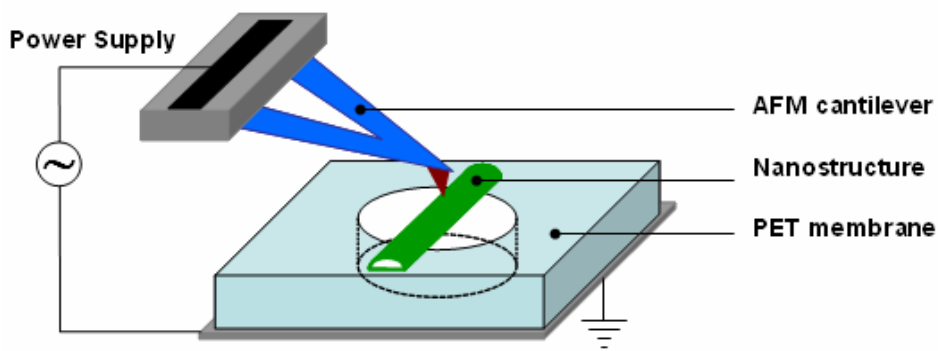

**Figure 4.**

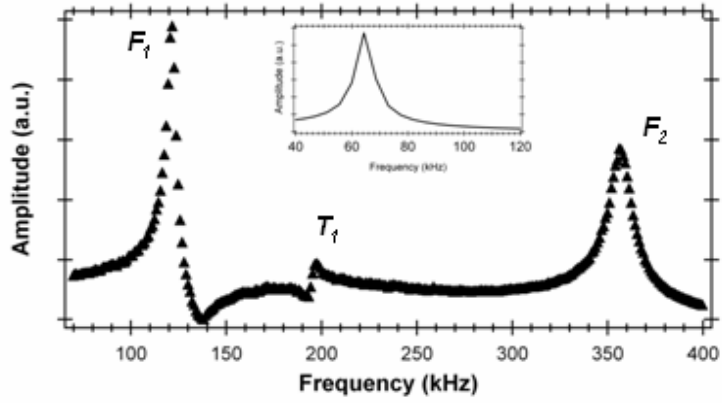

**Figure 5.**

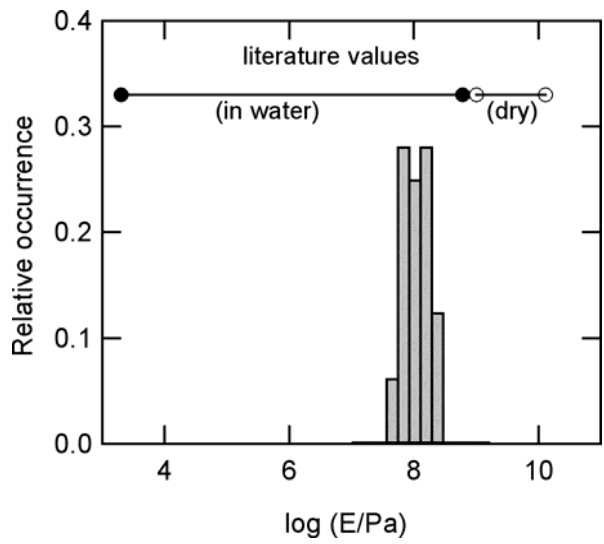

**Electronic-only material:**

**Computation of the moment of inertia of the LbL nanostructures.**

In our work, the elasticity modulus is determined using the classical formulae for the deflection of beams, without any assumptions on the contact mechanics (i.e., in the present case, on the tip-nanostructure contact). However, to obtain accurate values of Young's modulus, the moment of inertia of the probed nanostructures must be known, which requires knowing with precision their exact dimensions and shapes. Due to deformation upon adsorption on the membrane, the height measured from the cross-section of the AFM image of the nanostructure, $h_{AFM}$, is always lower than the pore diameter ($\phi_0$) used for the fabrication of the nanotubes. Therefore, the nanostructure was assumed to adopt the shape of a flat ribbon for its part in contact with the membrane, whereas its top part was considered to be a half-ellipsoid (figure 1).

Although no adhesion exists in the suspended region, the nanotubes may keep their flattened shape due to their softness. Therefore, a general model with an ellipsoid shape should be assumed. However, as the moments of inertia for the two shapes (the ellipsoid shape and our modelled shape) are of the same order of magnitude (the difference is lower than 5% for the probed nanostructures), we assumed that the same shape was conserved over the whole nanostructure length.

To compute the moment of inertia (I) of this model shape, precise geometrical parameters must be known for the nanostructures. First, the nanostructure width, W, which cannot be directly obtained from the AFM images due to tip dilation effects. Therefore, W was obtained from volume conservation between the deformed and undeformed shapes (nanotubes within pores vs flattened nanostructures on membranes), according to:

$$W = \frac{\pi(\phi_0 - h_{AFM}/2)}{1 + \pi/4}$$

It should be noted that other expressions could be taken, depending on the (unknown) details of the deformation of the nanotubes. However, not only the final value of W would not be seriously affected by these alternate choices [1-4], but this model shape seems particularly well adapted to the TEM and AFM images.

Second, the center of gravity (G) of the cross-section of the modeled shape was computed from the center of gravity of each component (ribbon part and half-ellipse part) and from their respective cross-section. The center of gravity is defined as the ratio between the moment about the x-axis and the mass of the structure. In the case of an empty half-ellipse, the center of gravity is expressed by:

$$y_{C-he} = \frac{\frac{4}{3}ab^2 - 2(b-t)^2 a + \frac{2}{3}\left(\frac{b-t}{a-t}\right)^2 a^3}{\pi(a+b-t)t}$$

where a represents half the nanostructure width (a = W/2), b is the nanostructure height minus the nanostructure thickness (b = $h_{AFM}$ -t). The nanostructure thickness, t, was obtained from TEM measurements of the pore sizes before and after LbL deposition, as described in ref [5].

From this relation, the center of gravity of each deformed nanostructure can be calculated as:

$$y_C = \frac{\frac{1}{2}\pi(a+b-t)t^2 - \frac{1}{3}ab^2 + 2abt + \frac{1}{3}\left(\frac{b-t}{a-t}\right)^2 a^3}{2at + \frac{1}{2}\pi(a+b-t)t}$$

Then, by applying Huygens' theorem, the moment of inertia about the x-axis of the whole modeled shape with respect to its own center of gravity ($y_c$) was determined. In this theorem, the moment of inertia of the ribbon part ($I_{rib}$) has to be known as well as the moment of inertia corresponding to the empty half-ellipse shape ($I_{he}$). For the first one, the moment of inertia is that of a classical rectangular beam :

$$I_{rib} = \frac{at^3}{6}$$

For the second one, the moment of inertia can be found by subtracting the moment of inertia of the "missing" inner half-ellipse from the outer one.

$$I_{he} = \frac{\pi}{8}\left[ab^3 - (a-t)(b-t)^3\right]$$

By this theorem, the moment of inertia of each probed nanostructure could be estimated from its measured geometrical dimensions. Given the assumptions made to compute the moment of inertia, we estimate the precision on its value to be on the order of 10%.

**Fig. 1.** Model of the shape of a polyelectrolyte nanostructure lying on a membrane.

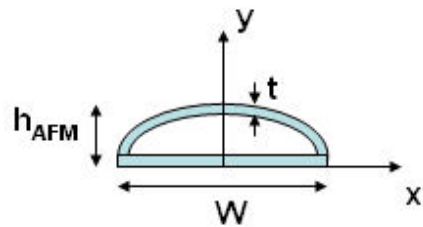

**Fig. 2.** Model shape of a polyelectrolyte nanostructure (deformed nanotube) lying on a membrane.

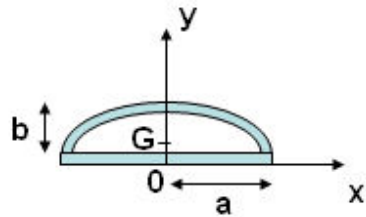